\documentclass[prb,amssymb,twocolumn,showpacs,superscriptaddress,floatfix]{revtex4}

\usepackage{graphicx}
\usepackage{amsmath}
\usepackage{bm}

\begin{document}

\title{Transverse rectification of disorder-induced fluctuations in a driven system
}

\author{Alejandro B. Kolton}
\affiliation{Universit\'e de Gen\`eve, DPMC, 24 Quai Ernest Ansermet,
CH-1211 Gen\`eve 4, Switzerland 
}

\thanks{\emph{Present address:} Dept. de F\'{\i}sica At\'omica, Molecular y Nuclear, 
Universidad Complutense de Madrid, 28040 Madrid, Spain.}

\begin{abstract}
We study 
numerically 
the overdamped 
motion of 
particles driven in 
a two dimensional ratchet potential. 
In the proposed design, of the so-called geometrical-ratchet type, 
the mean velocity of a single particle in response 
to a constant force has a transverse component that can be 
induced by the presence of thermal or other unbiased 
fluctuations.
We find that additional quenched disorder 
can strongly enhance the transverse drift at low temperatures, 
in spite of reducing the transverse mobility. We show that, under general conditions, 
the rectified transverse velocity of a driven particle fluid is 
equivalent to the response of a one dimensional flashing ratchet working at a 
drive-dependent effective temperature, defined through generalized 
Einstein relations. 
\end{abstract}


\maketitle 

The idea of generating  a directed dissipative transport 
in a system kept out of thermal equilibrium  
only by unbiased
perturbations 
has motivated an outburst of experimental and 
theoretical works in the last years. 
The ratchet effect 
is indeed of interest, both for applications 
and modelling, in very diverse systems, ranging from 
biological motors, 
\cite{molecular_motors_astumian_bier}  
colloidal matter, 
\cite{colloids_marquet} 
granular matter, 
\cite{granular_farkas}
vortex matter in superconductors,  
\cite{vortices_lee}
Josephson junction arrays, 
\cite{josephson_falo} 
atoms in optical traps, 
\cite{coldatoms_lundh} 
electrons in semiconductor heterostructures, \cite{electrons_linke} to gambling games. \cite{parrondo_paradox}
One of the simplest models is the ``flashing ratchet'', where a directed motion 
of a Brownian particle (i.e. breaking of the detailed balance condition) is obtained by 
coupling it to a pulsating asymmetric-periodic potential. The identification of the essential physical ingredients 
for the effect shows that a large variety 
of ratchets 
and rectification mechanisms 
can be realized. 
\cite{review_reimann} 
\begin{figure}
\centerline{\includegraphics[height=7.0cm]{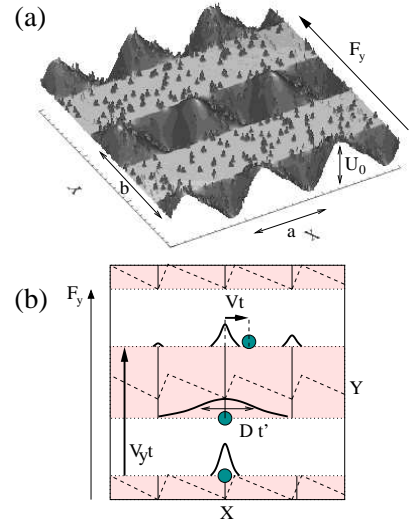}}
\caption{(a) Ratchet potential with disorder.
(b) Schematics of the transverse rectification mechanism (top view). Particles move 
with an average velocity $V_y$ in the direction of the applied force $F_y$. In 
the white regions the interaction with the (attractive or repulsive)  
centers and with the thermal bath induces diffusion 
in the non-driven direction 
In the shaded regions 
a periodic-asymmetric potential 
tends to localize particles at its minima. An average 
transverse shift (see circles) is produced at a rate $V$.
}
\label{schema}
\end{figure}

Recently, there has been a growing interest in the so-called {\it geometrical ratchets} since they can be used as 
continuous ``molecular sieves'' to separate particles experimentally (such as macromolecules or mesoscopic objects), according to its physical properties. 
These devices are typically two-dimensional systems containing a periodic array of asymmetric obstacles. By driving the particles 
through the array an average lateral drift appears, as transverse diffusive motion is rectified by the collisions with the asymmetric obstacles. 
Different types of geometrical ratchets have been analyzed in the literature, both experimentally \cite{Oudenaarden_geometrical_ratchet} and theoretically. 
\cite{Ertas_geometrical_ratchet}
The effect of additional quenched disorder in these two-dimensional systems has not been discussed yet, though interesting anomalous transport properties of one-dimensional disordered ratchet systems were reported. \cite{harms_disorder_ratchet_1d} 
Such a study is not only relevant for applications where disorder can not be avoided, but it is also 
an interesting and challenging issue. The 
driven motion of particles in a disordered substrate yields a non-trivial hydrodynamics. 
The current-driven motion of vortices in type II superconductors is a prominent example, where disorder, 
apart from reducing dissipation, is responsible for marked non-equilibrium transport and magnetic properties. \cite{giamarchi_vortex_review}
On the other hand, already the simpler case of driven non-interacting Brownian particles in two dimensions displays complex 
phenomena. While diffusion is anomalous at equilibrium,\cite{bouchaud} under a finite drive diffusion becomes normal 
in the co-moving frame, with anisotropic and velocity-dependent diffusion constants and mobilities.\cite{kolton_anisotropic_difusion} 
Moreover, a disordered substrate can provide alone a local or global ratchet effect, such as 
the generation of large-scale vorticity in the probability current by driving particles with an 
uniform alternate drive,\cite{makeev_large_scale_vorticity} and the net directed motion produced by 
driving the particles with crossed ac-drives. 
\cite{reichhardt_disorder_ratchet_2d}

In this paper we investigate the effect of quenched disorder in a simple geometrical ratchet design, under an uniform and constant 
driving force. We find that disorder can strongly enhance the transverse drift both for non-interacting and interacting particles, thus improving the performance of the device for applications. We show that the transverse velocity of a driven fluid is equivalent to the 
response of a one-dimensional flashing ratchet working at a drive-dependent effective temperature, defined through generalized fluctuation-dissipation relations.

Let us consider the overdamped motion of particles in a 
two dimensional potential like the one depicted in Fig.\ref{schema}. 
The equation of motion of a particle 
in position ${\bf R}_i$ is:
\begin{eqnarray}
\eta \frac{d{\bf R}_i}{dt} = -{\bm \nabla}_i \biggl[\sum_{j\not= i} V(R_{ij}) + 
U({\bf R}_i) \biggr]+ {\bf F} + 
{\bm \zeta}_i(t),
\label{eqmotion}
\end{eqnarray} 

where $R_{ij}=|{\bf R}_i-{\bf R}_j|$ is the distance between particles $i,j$,
$R_{ip}$ is the distance between the particle $i$ and
a site at ${\bf R}_p$, $\eta$ is the friction, 
and ${\bf F}=F_y {\hat y}$ is the driving force. The effect of a thermal bath 
at temperature $T$ is given by the stochastic
force ${\bm\zeta}_i(t)$,  satisfying $\langle {\zeta}^\mu _i(t) 
\rangle=0$ and $\langle {\zeta}^\mu   _i(t){\zeta}^{\mu '}_j(t') 
\rangle = 2 \eta k_B T \delta(t-t') \delta_{ij} \delta_{\mu \mu '}$, 
where  $\langle \ldots \rangle$ denotes average over the ensemble 
of ${\bm\zeta}_i$. For concreteness we consider a 
logarithmic repulsive particle-particle interaction $V(r)=-A_v\ln(r)$
which corresponds for instance to the vortex-vortex 
interaction in 2D thin film superconductors \cite{kolton_transverse_freezing}. 
Particles interact with the quenched potential
$U({\bf R})=U_R({\bf R})+ U_p({\bf R})$.  
$U_R$ is a ratchet potential with the form,
$U_R({\bf R})=\frac{a}{2\pi}F_R(Y)G_R(X)$, 
where $X \equiv {\bf R}.{\hat x},\;Y \equiv {\bf R}.{\hat y}$, 
$G_R(X)=\sin (2 \pi X/a)+0.25 \sin (4 \pi X/a)$ 
and $F_R(y)=U_0 \cos(2\pi y/b)\Theta[\cos(2\pi y/b)]$, with $\Theta$ 
the Heaviside function. This ratchet potential is similar to a 
periodic array of obstacles, asymmetrical around the $x-axis$ 
but symmetrical around the $y-axis$, as the ones considered in 
Ref..
\cite{Ertas_geometrical_ratchet} 
Disorder is short-range correlated, and it is modeled as a random 
distribution of centers such that 
$U_p({\bf R}_i) = \sum_p A_p e^{-(R_{ip}/r_p)^2}$, where 
$R_{ip}=|{\bf R}_i-{\bf R}_p|$ is the distance between the particle 
$i$ and a center at ${\bf R}_p$. Centers can be either 
attractive $A_p<0$ (wells) or repulsive $A_p>0$ (humps) or a 
combination of both. We solve Eq.\ref{eqmotion} numerically 
by using the method of Ref.\cite{kolton_teff}.
Length is normalized by $r_p$, energy by $2 A_p$, 
and time by  $\tau=\eta r_p^2/|2 A_p|$.  We consider $N=60$ 
particles and $N_p$ pinning centers in a rectangular box 
of size $L_x\times L_y$ and periodic boundary conditions, 
with $L_y=100$, $L_x=20\sqrt{3} L_y$. We average 
calculated properties over $500$ disorder realizations.

\begin{figure}
\centerline{\includegraphics[width=7.0cm]{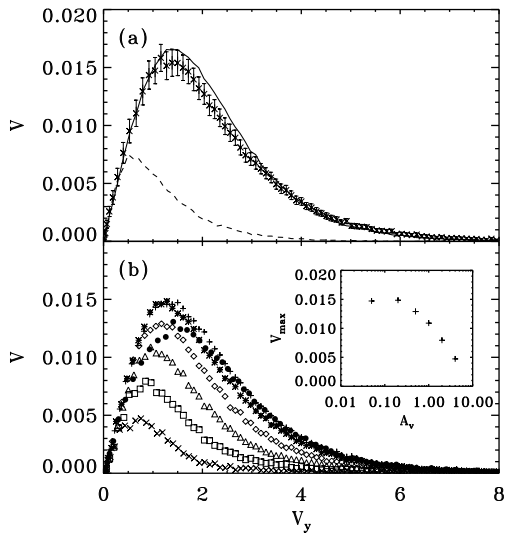}}
\caption{Transverse velocity $V$ vs longitudinal  
velocity $V_y$. (a) 
Symbols correspond to the disordered system at $T=0.05$, the dashed-line 
to the clean system at $T=0.05$ and the solid-line to the clean 
system at an effective temperature $T_{\tt eff}(V_y,T)$. (b) 
$V$ vs $V_y$ for different interaction strengths $A_v$: 
$A_v=0.05$ ($+$),  $A_v=0.2$ ($\ast$),  $A_v=0.5$ ($\diamond$),  
$A_v=1$ ($\vartriangle$), $A_v=2$ ($\square$), $A_v=5$ ($\times$). 
($\bullet$) symbols correspond to purely attractive 
pinning centers, with $A_p=-0.5$, at $T=0.05$. 
Inset: maximun rectification as a function of $A_v$.}
\label{recti}
\end{figure}

We start by discussing the simplest case of non-interacting particles, $A_v=0$, 
without disorder, $N_p=0$, and with a ratchet potential of amplitude 
$U_0=1$. The dashed-lines of Fig.\ref{recti}(a) show 
the transverse drift rate $V \equiv \langle \frac{1}{N} \sum_i \frac{dX_i}{dt} \rangle$ 
at $T=0.05$ as a function of the longitudinal 
velocity $V_y \equiv \langle \frac{1}{N} \sum_i \frac{dY_i}{dt} \rangle$. 
We see that the transverse velocity $V$ increases from zero, has a maximun 
$V \sim 0.0075$ at $V_y\sim 0.5$, and decays to zero at 
large longitudinal velocity.
Since $V=0$ at $T=0$, the average directed transverse 
motion observed is induced by the thermal noise. 
This rectification effect is easy to understand, and Fig.\ref{schema}(b) 
illustrates the mechanism, where the 
transverse diffusion constant is 
$D = 2T$ if there is no disorder nor inter-particle interactions. 
For our discussion it is useful to make explicit 
the connection between the type of response 
shown in Fig.\ref{recti}(a) and the one 
of a flashing ratchet. If $F_y$ is large, the 
mean velocity in the 
driven direction $V_y \equiv \langle \frac{dY}{dt} \rangle$ 
is $V_y \sim F_y - O(F_y^{-1})$ and 
longitudinal fluctuations are much smaller (by a factor $O(F_y^{-1})$)
than transverse fluctuations.\cite{kolton_anisotropic_difusion} 
At $T=0$, the equation of motion for the coordinate $X$ of a particle 
located at ${\bf R}=X {\hat x}+Y {\hat x}$ can be thus written as 
\begin{eqnarray}
\frac{d X}{dt} &\approx& -F_R(V_y t) G'_R(X).
\end{eqnarray}
Since $F_R(V_y t)=U_0 \cos(2\pi V_y t/b)\Theta[\cos(2\pi V_y t/b)]$,  
$X$ feels the ratchet potential $G_R(X)$ switching on and off 
periodically with time periods $\tau_{\tt on} = \tau_{\tt off} = b/2 V_y$. 
At small drives the mechanism is the same although $\tau_{\tt on}$ 
becomes increasingly larger than $\tau_{\tt off}$ since the 
wells of $U_R(X,Y)$ delay the motion in the $y$ direction.
(see Fig.\ref{schema}). The mapping to a flashing ratchet 
explains the observed directed transverse 
motion with $V>0$ when $T>0$ and can be thus used 
as an effective model to explain all the features of the 
response shown in Fig.\ref{recti}(a).

The effect discussed so far 
is similar to the one described in Ref.. 
\cite{Ertas_geometrical_ratchet} 
Let us 
now add disorder, by putting $N_p=2000$ randomly 
located pinning sites. The resulting response is 
shown in Fig.\ref{recti}(a) (symbols). As 
we can see, disorder strongly enhances the rectification at 
intermediate and large longitudinal velocity and also broaden 
the range of $V_y$ where response is appreciable, with respect 
to the clean case. In addition, we find that at intermediate and 
large $V_y$, $V$ is finite even in the $T=0$ limit, 
since disorder induces transverse diffusion when $V_y>0$, even in the 
absence of thermal fluctuations. 
Finally, let us now turn-on the repulsive interaction between particles. In 
Fig.\ref{recti}(b) we show $V$ as a 
function of $V_y$, for different values of the repulsion strength $A_v$. 
In the inset of Fig.\ref{recti}(b) we see that 
the maximun response $V_{max}$ is almost 
constant with $A_v$ up to values $A_v \sim 0.2$ where a slow decay starts, but it is 
larger than the response of the clean system up to $A_v=2$. 
The decay of the response at large $A_v$ is explained 
by the decrease of 
transverse wandering due to increasingly correlated collective motion.
\cite{kolton_transverse_freezing} In Fig.\ref{recti}(b) we show that 
the response for purely attractive pinning centers $A_p=-1$ is smaller than for 
repulsive centers $A_p>1$ 
for small values of $V_y$, but indistinguishable for larger values of $V_y$. This is due to the 
fact that, at the density of centers considered, attractive centers are more effective to pin particles than 
humps, since the latter can provide two-dimensional pinning only by forming rare geometric traps. However 
at a density $N_p/L_x L_y \sim 1/r_p^2$ all these differences disappear completely. 
\begin{figure}
 \centerline{\includegraphics[width=7.0cm]{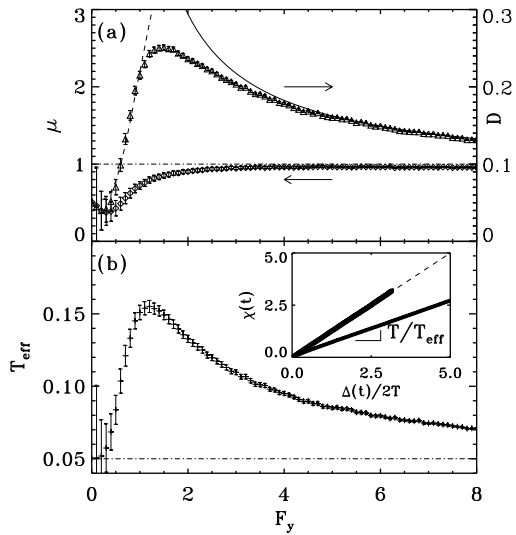}}
\caption{Motion in a purely disordered potential as function 
of $F_y$. 
(a) Transverse ($x$ direction) diffusion constant $D$ 
and mobility $\mu$. Dashed and solid lines 
indicate asymptotic forms of $D$. (b) 
Effective temperature $T_{\tt eff} \sim D/2\mu$. The dashed-dotted line indicates 
the thermal bath temperature, $T=0.05$. The inset shows that $T_{\tt eff}$ satisfies 
a generalized fluctuation-dissipation relation between 
the integrated response $\chi(t)$ and the quadratic mean 
displacement $\Delta(t)$. The dashed line corresponds to the equilibrium 
fluctuation-dissipation relation, which is valid only at short 
times $t \lesssim \xi/V_y$. Upper symbols correspond 
to $F_y=0.0$, and lower symbols to $F_y=4.0$.}
\label{vf}
\end{figure}

In order to understand the rectification characteristics described above it is 
instructive to study, separately, the motion of particles in the purely disordered case, without 
the ratchet potential (i.e. $U_0=0$). For simplicity we consider only 
the case of non-interacting particles, but we expect similar results for 
interacting particles in the dynamical regimes where transverse diffusion is non-zero. \cite{kolton_teff,kolton_unpublished} 
We analyze in detail the non-equilibrium transverse fluctuations as a function of $F_y$, since they 
affect directly the rectification in the presence of the ratchet potential. Following 
Ref.\cite{kolton_teff} we define the 
observables 
$O(t)=\frac{1}{N_v}\sum_{i=1}^{N_v} s_i  X_i(t)$ and
$\tilde{O}(t)=\sum_{i=1}^{N_v} s_i  X_i(t)$, 
where $s_i=-1, 1$ are random numbers with
$\overline{s_i}=0$ and $\overline{s_i s_j}=\delta_{ij}$.
The quadratic mean displacement can be written as 
$\Delta(t,t_0) \equiv \frac{1}{N_v}\sum_{i=1}^{N_v}
\langle|X_i(t)-X_i(t_0)|^2\rangle  
= C(t,t)+C(t_0,t_0)-2C(t,t_0)
$, with $C(t,t_0) = \overline{\langle O(t) \tilde{O}(t_0)\rangle}$.
The integrated response
function $\chi$ for the observable $O$ is obtained by applying
a perturbative force ${\bf f}_i = \epsilon s_i \hat{x}$
at time $t_0$ and keeping it constant for all subsequent times
on each particle, $
\chi(t,t_0) = \lim_{\epsilon \to 0}\frac{1}{\epsilon}
\Bigl[ \overline{\langle {O(t)} \rangle}_{\epsilon} -
\overline{\langle {O(t)} \rangle}_{\epsilon=0}
\Bigr]
$. In the steady state $\Delta(t,t_0)=\Delta(t-t_0)$ and $\chi(t,t_0)=\chi(t-t_0)$
and in particular at equilibrium the fluctuation-dissipation 
theorem (FDT) imposes $\chi(t)= \Delta(t)/2T$. When $F_y>0$  
the system is out of equilibrium and the FDT does not hold. We will 
show however, that generalized fluctuation-dissipation relations can still 
be defined for our system. In the long time limit we 
find $\Delta(t)\sim D t$ and $\chi(t)\sim \mu t$ thus allowing 
us to define the transverse diffusion constant $D$ and the transverse mobility $\mu$. These 
two quantities depend on the longitudinal driving force as shown in Fig.\ref{vf}(b). 
$D$ is non-monotonic, has a peak at $F_y \sim 1.5$, and 
decays approximately as a power law towards the equilibrium value without 
disorder $2T$ for large forces. This behavior can be 
understood by considering the effective transverse random walk  
induced only by the collisions with the pinning centers at $T=0$, and 
by simple heuristic arguments is possible to find the asymptotic forms $D \sim n_p r_p^3 V_y$ at small $V_y$ 
and $D \sim n_p r_p A_p^2/V_y$ at large $V_y$,\cite{kolton_anisotropic_difusion} indicated in Fig.\ref{vf}(b).  
At large $F_y$ the transverse mobility $\mu$ approaches the equilibrium value 
without disorder, $\mu = 1$ (independent of $T$). At small $F_y$, $\mu$ 
decreases due to trapping and its value at the limit $F_y \rightarrow 0$ is 
controlled by $T$. 
In the inset of Fig.\ref{vf}(c) we show the parametric plot 
of $\chi(t)$ vs  
$\Delta(t)/2T$ for $F_y=0$ (equilibrium) and $F_y=4.0$ (out of equilibrium). 
We see that the equilibrium FDT holds for $F_y=0$ as expected. 
For $F_y=4.0$ (and in general for $F_y>0$) we observe instead that the 
FDT holds only at very short time scales, 
$t \lesssim r_p/V_y$, but it is violated at 
long times. The type of violation observed can be quantified using 
the notion of time-scale dependent ``effective temperatures''  
introduced by Cugliandolo, Kurchan and Peliti.\cite{CKP} Following 
Ref.\cite{kolton_teff} we define a velocity-dependent transverse effective 
temperature $T_{\tt eff}$ from the slope shown in the inset of Fig.\ref{vf}(c). At long times 
this implies the generalized Einstein relation $T_{\tt eff} \sim D / 2\mu$ in the 
non-driven direction. In Fig.\ref{vf}(c) we see that $T_{\tt eff}$ follows 
closely $D$ except at low forces where $\mu$ decreases towards the value $\mu \sim 0.5$ 
at very low forces. For interacting particles 
similar results for $T_{\tt eff}$ were 
obtained at low and intermediate forces, in the plastic and smectic regimes 
of motion.\cite{kolton_transverse_freezing} At large forces and small 
temperatures however, the formation of 
pinned rough channels for particle motion 
\cite{giamarchi_vortex_review} leads to a transverse freezing 
of the moving vortex fluid \cite{kolton_transverse_freezing} at $T=0$. 
At this dynamical transition $D$ and $\mu$ are strongly reduced, and vanish 
at $T=0$ strictly.\cite{kolton_teff,kolton_anisotropic_difusion}

The fluctuating motion in a purely disordered substrate and the 
connection between the response of a geometrical and a flashing 
ratchet discussed above motivate a simple model that can be used 
to describe the rectification response of the 
disordered geometrical ratchet. We propose the 
following equation for the transverse 
motion of non-interacting particles,
\begin{eqnarray}
\frac{1}{\mu(V_y)}  \frac{d X_i}{dt} \approx  
 -F_R(V_y t) G'_R(X_i) + {\zeta^i_{\tt eff}}(t). 
\label{effe}
\end{eqnarray}
where we have replaced $Y_i \sim V_y t$ as before, the pinning force 
by an effective thermal noise ${\zeta^i_{\tt eff}}(t)$ and the bare 
mobility by $\mu(V_y)$. We use 
that $\langle {\zeta^i_{\tt eff}}(t) 
\rangle=0$ and $\langle {\zeta^i_{\tt eff}}(t){\zeta^{j}_{\tt eff}}(t') 
\rangle = (2 T_{\tt eff}(V_y)/\mu(V_y)) \delta(t-t') \delta_{ij}$ where 
$T_{\tt eff}(T,V_y)$ and $\mu(T,V_y)$ are the ones shown in Fig.\ref{vf}(a). 
Eq. \ref{effe} therefore models transverse motion in a 
coarse-grained way (in time and space), satisfying the generalized 
fluctuation-dissipation relation shown in the inset of 
Fig.\ref{vf}(b) (in the absence of the
ratchet potential). 
By construction, the assumptions of the 
model are that (i) transverse forces are small compared with the 
longitudinal drive $F_y$ and (ii), the particle motion is incoherent 
at the length scales of the ratchet potential. In Fig.\ref{recti}(a)
we see that the transverse drift generated by this model is 
close to the one of the full model, for the 
parameters analyzed in this paper. The rectification 
characteristics of the two-dimensional geometrical 
ratchet are therefore well described by the one-dimensional 
flashing-ratchet described by Eq.\ref{effe} 
working at the effective temperature $T_{\tt eff}(T,V_y)$ and 
friction $\mu^{-1}(T,V_y)$, determined by the 
disorder and the longitudinal velocity. Using this model 
the enhancement of the rectification observed in 
Fig.\ref{recti}(a) can be simply attributed to the fact 
that $T_{\tt eff} > T$, but with $T_{\tt eff}$ still 
smaller than the optimal temperature for rectification 
of the effective pulsating ratchet 
Eq.\ref{effe} 
is also expected to work for interacting particles by 
using the respective $T_{\tt eff}(T,V_y)$ and $\mu(T,V_y)$
except at low $T$ and large $F_y$ where condition (ii) can 
be violated since particles become correlated over long 
times and distances.\cite{kolton_unpublished}

In conclusion, we have studied numerically the effect 
of quenched disorder in a geometrical ratchet. We find 
that disorder enhances the transverse rectified 
velocity of a driven fluid. If particle motion is 
incoherent at the scale of the ratchet potential 
the response can be simply described by a 
one-dimensional flashing ratchet working at a 
disorder-induced, drive-dependent effective mobility 
and temperature, satisfying generalized Einstein relations. 
This effect can be used experimentally to enhance and control the 
performance of geometrical ratchets.

We acknowledge 
discussions with L.F. Cugliandolo, 
D. Dom\'{\i}nguez, T. Giamarchi, V.I. Marconi, 
and A. Rosso. This work was supported 
in part by the Swiss National Fund under Division II.


\begin{thebibliography}{52}
\expandafter\ifx\csname natexlab\endcsname\relax\def\natexlab#1{#1}\fi
\expandafter\ifx\csname bibnamefont\endcsname\relax
  \def\bibnamefont#1{#1}\fi
\expandafter\ifx\csname bibfnamefont\endcsname\relax
  \def\bibfnamefont#1{#1}\fi
\expandafter\ifx\csname citenamefont\endcsname\relax
  \def\citenamefont#1{#1}\fi
\expandafter\ifx\csname url\endcsname\relax
  \def\url#1{\texttt{#1}}\fi
\expandafter\ifx\csname urlprefix\endcsname\relax\def\urlprefix{URL }\fi
\providecommand{\bibinfo}[2]{#2}
\providecommand{\eprint}[2][]{\url{#2}}

\bibitem[{\citenamefont{Astumian and
  Bier}(1994)}]{molecular_motors_astumian_bier}
\bibinfo{author}{\bibfnamefont{R.~D.} \bibnamefont{Astumian}} \bibnamefont{and}
  \bibinfo{author}{\bibfnamefont{M.}~\bibnamefont{Bier}},
  \bibinfo{journal}{Phys. Rev. Lett.} \textbf{\bibinfo{volume}{72}},
  \bibinfo{pages}{1766} (\bibinfo{year}{1994});
\bibinfo{author}{\bibfnamefont{R.~D.} \bibnamefont{Astumian}},
  \bibinfo{journal}{Science} \textbf{\bibinfo{volume}{276}},
  \bibinfo{pages}{917} (\bibinfo{year}{1997});
\bibinfo{author}{\bibnamefont{J\"{u}licher}},
  \bibinfo{author}{\bibfnamefont{A.}~\bibnamefont{Adjari}}, \bibnamefont{and}
  \bibinfo{author}{\bibfnamefont{J.}~\bibnamefont{Prost}},
  \bibinfo{journal}{Rev. Mod. Phys.} \textbf{\bibinfo{volume}{69}},
  \bibinfo{pages}{1269} (\bibinfo{year}{1997}).

\bibitem[{\citenamefont{Marquet et~al.}(2002)\citenamefont{Marquet, Buguin,
  Talini, and Silberzan}}]{colloids_marquet}
\bibinfo{author}{\bibfnamefont{C.}~\bibnamefont{Marquet} {\it et~al.}},
  \bibinfo{journal}{Phys. Rev. Lett.} \textbf{\bibinfo{volume}{88}},
  \bibinfo{pages}{168301} (\bibinfo{year}{2002});
\bibinfo{author}{\bibfnamefont{S.~H.} \bibnamefont{Lee} {\it et~al.}},
  \bibinfo{journal}{Phys. Rev. Lett.} \textbf{\bibinfo{volume}{94}},
  \bibinfo{pages}{110601} (\bibinfo{year}{2005});
\bibinfo{author}{\bibfnamefont{D.}~\bibnamefont{Babic}} \bibnamefont{and}
  \bibinfo{author}{\bibfnamefont{C.}~\bibnamefont{Bechinger}},
  \bibinfo{journal}{Phys. Rev. Lett.} \textbf{\bibinfo{volume}{94}},
  \bibinfo{pages}{148303} (\bibinfo{year}{2005});
\bibinfo{author}{\bibfnamefont{A.}~\bibnamefont{Lib\'al} {\it et~al.}},
  \bibinfo{journal}{Phys. Rev. Lett.} \textbf{\bibinfo{volume}{96}},
  \bibinfo{pages}{188301} (\bibinfo{year}{2006}).

\bibitem[{\citenamefont{Farkas et~al.}(1999)\citenamefont{Farkas, Tegzes,
  Vukics, and Vicsek}}]{granular_farkas}
\bibinfo{author}{\bibfnamefont{Z.}~\bibnamefont{Farkas} {\it et~al.}},
  \bibinfo{journal}{Phys. Rev. E} \textbf{\bibinfo{volume}{60}},
  \bibinfo{pages}{7022} (\bibinfo{year}{1999});
\bibinfo{author}{\bibfnamefont{J.~F.} \bibnamefont{Wambaugh}},
  \bibinfo{author}{\bibfnamefont{C.}~\bibnamefont{Reichhardt}},
  \bibnamefont{and} \bibinfo{author}{\bibfnamefont{C.~J.} \bibnamefont{Olson}},
  \bibinfo{journal}{Phys. Rev. E} \textbf{\bibinfo{volume}{65}},
  \bibinfo{pages}{031308} (\bibinfo{year}{2002});
\bibinfo{author}{\bibfnamefont{D.}~\bibnamefont{{van der Meer}} {\it et~al.}},
  \bibinfo{journal}{Phys. Rev. Lett.} \textbf{\bibinfo{volume}{92}},
  \bibinfo{pages}{184301} (\bibinfo{year}{2004}).

\bibitem[{\citenamefont{Lee et~al.}(1999)\citenamefont{Lee, Jank\'o, Derenyi,
  and Barabasi}}]{vortices_lee}
\bibinfo{author}{\bibfnamefont{C.~S.} \bibnamefont{Lee} {\it et~al.}},
  \bibinfo{journal}{Nature (London)} \textbf{\bibinfo{volume}{400}},
  \bibinfo{pages}{337} (\bibinfo{year}{1999});
\bibinfo{author}{\bibfnamefont{J.~F.} \bibnamefont{Wambaugh} {\it et~al.}},
  \bibinfo{journal}{Phys. Rev. Lett.} \textbf{\bibinfo{volume}{83}},
  \bibinfo{pages}{5106} (\bibinfo{year}{1999});
\bibinfo{author}{\bibfnamefont{M.~B.} \bibnamefont{Hastings}},
  \bibinfo{author}{\bibfnamefont{C.~J.} \bibnamefont{Olson~Reichhardt}},
  \bibnamefont{and}
  \bibinfo{author}{\bibfnamefont{C.}~\bibnamefont{Reichhardt}},
  \bibinfo{journal}{Phys. Rev. Lett.} \textbf{\bibinfo{volume}{90}},
  \bibinfo{pages}{247004} (\bibinfo{year}{2003});
\bibinfo{author}{\bibfnamefont{C.~J.} \bibnamefont{Olson} {\it et~al.}},
  \bibinfo{journal}{Phys. Rev. Lett.} \textbf{\bibinfo{volume}{87}},
  \bibinfo{pages}{177002} (\bibinfo{year}{2001});
\bibinfo{author}{\bibfnamefont{B.~Y.} \bibnamefont{Zhu} {\it et~al.}},
  \bibinfo{journal}{Phys. Rev. B} \textbf{\bibinfo{volume}{68}},
  \bibinfo{pages}{014514} (\bibinfo{year}{2003});
\bibinfo{author}{\bibfnamefont{R.}~\bibnamefont{W\"{o}rdenweber}},
  \bibinfo{author}{\bibfnamefont{P.}~\bibnamefont{Dymashevski}},
  \bibnamefont{and} \bibinfo{author}{\bibfnamefont{V.~R.} \bibnamefont{Misko}},
  \bibinfo{journal}{Phys. Rev. B} \textbf{\bibinfo{volume}{69}},
  \bibinfo{pages}{184504} (\bibinfo{year}{2004});
\bibinfo{author}{\bibfnamefont{J.~E.} \bibnamefont{Villegas} {\it et~al.}},
  \textbf{\bibinfo{volume}{71}}, \bibinfo{pages}{024519}
  (\bibinfo{year}{2005}).

\bibitem[{\citenamefont{Falo et~al.}(1999)\citenamefont{Falo, Martinez, Mazo,
  and Cilla}}]{josephson_falo}
\bibinfo{author}{\bibfnamefont{F.}~\bibnamefont{Falo} {\it et~al.}},
  \bibinfo{journal}{Europhys. Lett.} \textbf{\bibinfo{volume}{45}},
  \bibinfo{pages}{024519} (\bibinfo{year}{1999});
\bibinfo{author}{\bibfnamefont{E.}~\bibnamefont{Trias} {\it et~al.}},
  \bibinfo{journal}{Phys. Rev. E} \textbf{\bibinfo{volume}{61}},
  \bibinfo{pages}{2257} (\bibinfo{year}{2000});
\bibinfo{author}{\bibfnamefont{G.}~\bibnamefont{Carapella}} \bibnamefont{and}
  \bibinfo{author}{\bibfnamefont{G.}~\bibnamefont{Costabile}},
  \bibinfo{journal}{Phys. Rev. Lett.} \textbf{\bibinfo{volume}{87}},
  \bibinfo{pages}{077002} (\bibinfo{year}{2001});
\bibinfo{author}{\bibfnamefont{D.~E.} \bibnamefont{Shal\'om}} \bibnamefont{and}
  \bibinfo{author}{\bibfnamefont{H.}~\bibnamefont{Pastoriza}},
  \bibinfo{journal}{Phys. Rev. Lett.} \textbf{\bibinfo{volume}{94}},
  \bibinfo{pages}{177001} (\bibinfo{year}{2005});
\bibinfo{author}{\bibfnamefont{V.~I.} \bibnamefont{Marconi}},
  \bibinfo{journal}{Physica C} \textbf{\bibinfo{volume}{437}},
  \bibinfo{pages}{195} (\bibinfo{year}{2006});
\bibinfo{author}{\bibfnamefont{A.}~\bibnamefont{Sterck}},
  \bibinfo{author}{\bibfnamefont{R.}~\bibnamefont{Kleiner}}, \bibnamefont{and}
  \bibinfo{author}{\bibfnamefont{D.}~\bibnamefont{Koelle}},
  \bibinfo{journal}{Phys. Rev. Lett.} \textbf{\bibinfo{volume}{95}},
  \bibinfo{pages}{177006} (\bibinfo{year}{2005}).

\bibitem[{\citenamefont{Lundh and Wallin}(2005)}]{coldatoms_lundh}
\bibinfo{author}{\bibfnamefont{E.}~\bibnamefont{Lundh}} \bibnamefont{and}
  \bibinfo{author}{\bibfnamefont{M.}~\bibnamefont{Wallin}},
  \bibinfo{journal}{Phys. Rev. Lett.} \textbf{\bibinfo{volume}{94}},
  \bibinfo{pages}{110603} (\bibinfo{year}{2005});
\bibinfo{author}{\bibfnamefont{R.}~\bibnamefont{Gommers}},
  \bibinfo{author}{\bibfnamefont{S.}~\bibnamefont{Bergamini}},
  \bibnamefont{and} \bibinfo{author}{\bibfnamefont{F.}~\bibnamefont{Renzoni}},
  \bibinfo{journal}{Phys. Rev. Lett.} \textbf{\bibinfo{volume}{95}},
  \bibinfo{pages}{073003} (\bibinfo{year}{2005});
\bibinfo{author}{\bibfnamefont{M.}~\bibnamefont{Schiavoni} {\it et~al.}},
  \bibinfo{journal}{Phys. Rev. Lett.} \textbf{\bibinfo{volume}{90}},
  \bibinfo{pages}{094101} (\bibinfo{year}{2003});
\bibinfo{author}{\bibfnamefont{R.}~\bibnamefont{Gommers}},
  \bibinfo{author}{\bibfnamefont{S.}~\bibnamefont{Denisov}}, \bibnamefont{and}
  \bibinfo{author}{\bibfnamefont{F.}~\bibnamefont{Renzoni}},
  \bibinfo{journal}{Phys. Rev. Lett.} \textbf{\bibinfo{volume}{96}},
  \bibinfo{pages}{240604} (\bibinfo{year}{2006}).


\bibitem[{\citenamefont{Linke et~al.}(2003)\citenamefont{Linke, Humphrey,
  Lofgren, Sushkov, Newbury, Taylor, and Omling}}]{electrons_linke}
\bibinfo{author}{\bibfnamefont{H.}~\bibnamefont{Linke} {\it et~al.}},
  \bibinfo{journal}{Science} \textbf{\bibinfo{volume}{286}},
  \bibinfo{pages}{2314} (\bibinfo{year}{2003}).

\bibitem[{\citenamefont{Parrondo and Din\'{\i}s}(2004)}]{parrondo_paradox}
\bibinfo{author}{\bibfnamefont{J.~M.~R.} \bibnamefont{Parrondo}}
  \bibnamefont{and}
  \bibinfo{author}{\bibfnamefont{L.}~\bibnamefont{Din\'{\i}s}},
  \bibinfo{journal}{Contemporary Physics} \textbf{\bibinfo{volume}{45}},
  \bibinfo{pages}{147} (\bibinfo{year}{2004}).

\bibitem[{\citenamefont{Reimann}(2002)}]{review_reimann}
\bibinfo{author}{\bibfnamefont{P.}~\bibnamefont{Reimann}},
  \bibinfo{journal}{Phys. Rep.} \textbf{\bibinfo{volume}{361}},
  \bibinfo{pages}{57} (\bibinfo{year}{2002});
\bibinfo{author}{\bibfnamefont{R.~D.} \bibnamefont{Astumian}} \bibnamefont{and}
  \bibinfo{author}{\bibfnamefont{P.}~\bibnamefont{H{\"{a}}nggi}},
  \bibinfo{journal}{Physics Today} \textbf{\bibinfo{volume}{55}},
  \bibinfo{pages}{33} (\bibinfo{year}{2002}).

\bibitem{Oudenaarden_geometrical_ratchet}
\bibinfo{author}{\bibfnamefont{A.} \bibnamefont{van Oudenaarden}} \bibnamefont{and}
  \bibinfo{author}{\bibfnamefont{S.~G.} \bibnamefont{Boxer}},
  \bibinfo{journal}{Science} \textbf{\bibinfo{volume}{285}},
  \bibinfo{pages}{1046} (\bibinfo{year}{1999}).

\bibitem[{\citenamefont{Erta\ifmmode~\mbox{\c{s}}\else
  \c{s}\fi{}}(1998)}]{Ertas_geometrical_ratchet}
\bibinfo{author}{\bibfnamefont{D.}~\bibnamefont{Ertas}}, \bibinfo{journal}{Phys. Rev. Lett.}
  \textbf{\bibinfo{volume}{80}}, \bibinfo{pages}{1548} (\bibinfo{year}{1998});
\bibinfo{author}{\bibfnamefont{T.~A.~J.} \bibnamefont{Duke}} \bibnamefont{and}
  \bibinfo{author}{\bibfnamefont{R.~H.} \bibnamefont{Austin}},
  \bibinfo{journal}{Phys. Rev. Lett.} \textbf{\bibinfo{volume}{80}},
  \bibinfo{pages}{1552} (\bibinfo{year}{1998});
\bibinfo{author}{\bibfnamefont{I.}~\bibnamefont{Der\'enyi}} \bibnamefont{and}
  \bibinfo{author}{\bibfnamefont{R.D.}~\bibnamefont{Astumian}},
  \bibinfo{journal}{Phys. Rev. E} \textbf{\bibinfo{volume}{58}},
  \bibinfo{pages}{7781} (\bibinfo{year}{1998});
\bibinfo{author}{\bibfnamefont{M.}~\bibnamefont{Bier} {\it et~al.}},
  \bibinfo{author}{\bibfnamefont{M.}~\bibnamefont{Kostur}},
  \textbf{\bibinfo{volume}{61}}, \bibinfo{pages}{7184} (\bibinfo{year}{2000});
\bibinfo{author}{\bibfnamefont{M.}~\bibnamefont{Kostur}} \bibnamefont{and}
  \bibinfo{author}{\bibfnamefont{L.}~\bibnamefont{Shimansky-Geier}},
  \bibinfo{journal}{Phys. Lett. A} \textbf{\bibinfo{volume}{265}},
  \bibinfo{pages}{337} (\bibinfo{year}{2000});
\bibinfo{author}{\bibfnamefont{C.}~\bibnamefont{Keller}},
  \bibinfo{author}{\bibfnamefont{F.}~\bibnamefont{Marquardt}},
  \bibnamefont{and} \bibinfo{author}{\bibfnamefont{C.}~\bibnamefont{Bruder}},
  \bibinfo{journal}{Phys. Rev. E} \textbf{\bibinfo{volume}{65}},
  \bibinfo{pages}{041927} (\bibinfo{year}{2002});
\bibinfo{author}{\bibfnamefont{S.}~\bibnamefont{Savel'ev} {\it et~al.}},
  \bibinfo{journal}{Phys. Rev. B} \textbf{\bibinfo{volume}{71}},
  \bibinfo{pages}{214303} (\bibinfo{year}{2005}).

\bibitem[{\citenamefont{Harms and Lipowsky}(1997)}]{harms_disorder_ratchet_1d}
\bibinfo{author}{\bibfnamefont{T.}~\bibnamefont{Harms}} \bibnamefont{and}
  \bibinfo{author}{\bibfnamefont{R.}~\bibnamefont{Lipowsky}},
  \bibinfo{journal}{Phys. Rev. Lett.} \textbf{\bibinfo{volume}{79}},
  \bibinfo{pages}{2895} (\bibinfo{year}{1997});
\bibinfo{author}{\bibfnamefont{F.}~\bibnamefont{Marchesoni}},
  \bibinfo{journal}{Phys. Rev. E} \textbf{\bibinfo{volume}{56}},
  \bibinfo{pages}{2492} (\bibinfo{year}{1997});
\bibinfo{author}{\bibfnamefont{M.~N.} \bibnamefont{Popescu} {\it et~al.}},
  \bibinfo{journal}{Phys. Rev. Lett.} \textbf{\bibinfo{volume}{85}},
  \bibinfo{pages}{3321} (\bibinfo{year}{2000}).

\bibitem[{\citenamefont{Giamarchi and
  Bhattacharya}(2002)}]{giamarchi_vortex_review}
\bibinfo{author}{\bibfnamefont{T.}~\bibnamefont{Giamarchi}} \bibnamefont{and}
  \bibinfo{author}{\bibfnamefont{S.}~\bibnamefont{Bhattacharya}}, in
  \emph{\bibinfo{booktitle}{High Magnetic Fields: Applications in Condensed
  Matter Physics and Spectroscopy}}, edited by
  \bibinfo{editor}{\bibfnamefont{C.}~\bibnamefont{{Berthier {\it et al.}}}}
  (\bibinfo{publisher}{Springer-Verlag}, \bibinfo{address}{Berlin},
  \bibinfo{year}{2002}), p. \bibinfo{pages}{314},
  \bibinfo{note}{cond-mat/0111052}.

\bibitem[{\citenamefont{Bouchaud and Georges}(1990)}]{bouchaud}
\bibinfo{author}{\bibfnamefont{J.-P.} \bibnamefont{Bouchaud}} \bibnamefont{and}
  \bibinfo{author}{\bibfnamefont{A.}~\bibnamefont{Georges}},
  \bibinfo{journal}{Phys. Rep} \textbf{\bibinfo{volume}{195}},
  \bibinfo{pages}{127} (\bibinfo{year}{1990}).

\bibitem[{\citenamefont{Kolton}(2006)}]{kolton_anisotropic_difusion}
\bibinfo{author}{\bibfnamefont{A.~B.} \bibnamefont{Kolton}},
  \bibinfo{journal}{Physica C} \textbf{\bibinfo{volume}{437}},
  \bibinfo{eid}{026112} (\bibinfo{year}{2006}).

\bibitem[{\citenamefont{Makeev et~al.}(2005)\citenamefont{Makeev, Derenyi, and
  Barabasi}}]{makeev_large_scale_vorticity}
\bibinfo{author}{\bibfnamefont{M.~A.} \bibnamefont{Makeev}},
  \bibinfo{author}{\bibfnamefont{I.}~\bibnamefont{Derenyi}}, \bibnamefont{and}
  \bibinfo{author}{\bibfnamefont{A.-L.} \bibnamefont{Barabasi}},
  \bibinfo{journal}{Phys. Rev. E} \textbf{\bibinfo{volume}{71}},
  \bibinfo{pages}{026112} (\bibinfo{year}{2005}).

\bibitem[{\citenamefont{Reichhardt and
  Olson~Reichhardt}(2006)}]{reichhardt_disorder_ratchet_2d}
\bibinfo{author}{\bibfnamefont{C.}~\bibnamefont{Reichhardt}} \bibnamefont{and}
  \bibinfo{author}{\bibfnamefont{C.~J.~Olson} \bibnamefont{Reichhardt}},
  \bibinfo{journal}{Phys. Rev. E} \textbf{\bibinfo{volume}{73}},
  \bibinfo{pages}{011102} (\bibinfo{year}{2006}).

\bibitem[{\citenamefont{Kolton et~al.}(2002)\citenamefont{Kolton, Exartier,
  Cugliandolo, Dom\'inguez, and Gr\o{}nbech-Jensen}}]{kolton_teff}
\bibinfo{author}{\bibfnamefont{A.~B.} \bibnamefont{Kolton} {\it et~al.}},
  \bibinfo{journal}{Phys. Rev. Lett.} \textbf{\bibinfo{volume}{89}},
  \bibinfo{pages}{227001} (\bibinfo{year}{2002}).


\bibitem[{\citenamefont{Kolton et~al.}(1999)\citenamefont{Kolton, Dom\'inguez,
  and Gr\o{}nbech-Jensen}}]{kolton_transverse_freezing}
\bibinfo{author}{\bibfnamefont{A.~B.} \bibnamefont{Kolton}},
  \bibinfo{author}{\bibfnamefont{D.}~\bibnamefont{Dom\'inguez}},
  \bibnamefont{and}
  \bibinfo{author}{\bibfnamefont{N.}~\bibnamefont{Gr\o{}nbech-Jensen}},
  \bibinfo{journal}{Phys. Rev. Lett.} \textbf{\bibinfo{volume}{83}},
  \bibinfo{pages}{3061} (\bibinfo{year}{1999}).

\bibitem[{\citenamefont{Kolton}()}]{kolton_unpublished}
\bibinfo{author}{\bibfnamefont{A.~B.} \bibnamefont{Kolton}}, \bibinfo{note}{to
  be published}.

\bibitem[{\citenamefont{Cugliandolo et~al.}(1997)\citenamefont{Cugliandolo,
  Kurchan, and Peliti}}]{CKP}
\bibinfo{author}{\bibfnamefont{L.~F.} \bibnamefont{Cugliandolo}},
  \bibinfo{author}{\bibfnamefont{J.}~\bibnamefont{Kurchan}}, \bibnamefont{and}
  \bibinfo{author}{\bibfnamefont{L.}~\bibnamefont{Peliti}},
  \bibinfo{journal}{Phys. Rev. E} \textbf{\bibinfo{volume}{55}},
  \bibinfo{pages}{3898} (\bibinfo{year}{1997}).

\end{thebibliography}

\end{document}